\documentstyle[12pt]{article}
\def\nue{{\nu_e}}
\def\anue{{\bar\nu_e}}
\def\numu{{\nu_{\mu}}}
\def\anumu{\bar{\nu_{\mu}}}
\def\nutau{{\nu_{\tau}}}
\def\anutau{{\bar\nu_{\tau}}}
\begin{document}

\begin{center}
{\large{ \bf { Constraints on Neutrino Mixing from r-process 
Nucleosynthesis in Supernovae}}} \\
\vskip 30pt
{\it  Sandhya Choubey $^{a}$ and Srubabati Goswami $^{b}$ \\
$^{a}$ Saha Institute of Nuclear Physics,\\1/AF, Bidhannagar,
Calcutta 700064, INDIA.\\
$^{b}$ Physical Research Laboratory, Ahmedabad 380009, INDIA}
\vskip 30pt

\bf{Abstract}
\end{center} 

{\small
In this paper we note that for neutrino 
mass squared differences in the 
range $10^{-4} - 10^{-1} eV^2$,
vacuum neutrino oscillations can take place between the neutrino-sphere 
and the weak freeze-out radius in a type II supernova. 
Requiring that such oscillations are consistent with the r-process
nucleosynthesis from supernovae one can constrain the mixing of $\nu_e$
with $\nu_{\mu}$ ( or $\nu_\tau$) down to $10^{-4} eV^2$. 
We first do a two-flavor study and find that the 
neutron rich condition $Y_e < 0.5$ is
satisfied for all values of mixing angles. However if we take the
criterion $Y_e < 0.45$  
for a successful r-process and assume $\nu_\mu - \nu_e$ oscillations to
be operative then this    
mode as a possible solution 
to the atmospheric neutrino anomaly is ruled out in accordance
with the recent CHOOZ result. Furthermore since we can probe 
mass ranges lower than CHOOZ 
the narrow range that was allowed by the CHOOZ
data at 99\% C.L. is also ruled out. 
Next we do a three-generation analysis keeping $\Delta m_{12}^2 \sim
10^{-5} 
eV^2$ or $10^{-11} eV^2$  (solar neutrino range) and 
$\Delta m_{13}^2 \approx \Delta m_{23}^2 \sim  10^{-4} - 10^{-1} eV^2$
(atmospheric 
neutrino range) 
and use the condition $Y_e < 0.45$  to give bounds on the mixing parameters
and compare our results with the CHOOZ bound. 
We also calculate the increase in the shock-reheating obtained 
by such oscillations.} 

\section{ Introduction}

Heavy neutron rich nuclei beyond the iron group are predominantly 
made by the rapid neutron 
capture process or the r-process. Over the last four  decades different 
astrophysical 
environments have been proposed as possible sites for r-process
nucleosynthesis 
\cite{one}. 
The very high neutron number densities $> 10^{20} cm^{-3}$,  
temperatures $\sim 2-3\times{10^9}^0 K $ \cite{four} and time scales 
$\sim 1s$, prompted 
investigators to suggest supernova as a plausible site for 
r-process nucleosynthesis \cite{five}. 
But where exactly in the supernova does the r-process actually take place is 
a question which is still to be answered. A putative r-process
 site suggested in 
the recent years is the neutrino heated ejecta from the post core bounce 
environment of a type II supernova or the "hot bubble" \cite{six,ww}.  
The major advantage which the "hot bubble" has over other proposed sites 
is that it correctly predicts that only $10^{-4} M_\odot$ of r-process 
nuclei are ejected per supernova \cite{mc}.
 The conditions of high temperatures and low density 
in the "hot bubble" makes it conducive for synthesising the right amount of 
r-process nuclei. 

Near the neutrino-sphere the temperature is high 
so that nuclear statistical equilibrum persists
and the mass fraction of the nuclei are determined 
by the Saha equation. 
Since the photon to baryon ratio is very high
in the "hot bubble", at equilibrium we have nucleons and alpha
particles as the most abundant species. It is on these nucleons 
that the $\nue$ and $\anue$ capture takes place which determines the
electron fraction $Y_e$. As the temperature drops below about 0.5 MeV, 
the expansion rate becomes faster than the nuclear reaction
rates and  one has the nuclear freeze-out. 
Below about 0.26 MeV the charge particle reactions
freeze and beyond this point temperatures become so low that one can
have only neutron captures on the heavy seed nuclei or the r-process.

For r-process to be possible in the supernova, the conditions should be 
neutron rich. The parameter which determines this is the electron
fraction $Y_e$ 
at the radius where the absorption of $\nue$ and $\anue$ on 
free nucleons freeze out. This is called the weak freeze-out radius 
($r_{\small {WFO}}$) and is found to very close to the nuclear 
freeze-out radius in most supernova models.
Since $\langle{E_{\nue}}\rangle~ < ~ \langle {E_{\anue}} \rangle~ 
<~\langle{E_{\numu}}\rangle$ 
($ \langle{E_{\nue}}\rangle , \langle{E_{\anue}}\rangle
~{\rm and}~ \langle{E_{\numu}}\rangle$ 
are the average energies of the $\nue, \anue$ and $\numu$ respectively 
and the energy spectrum of the $\numu, \anumu, \nutau$ and $\anutau$ are 
identical) and since 
$Y_e$ is determined by the properties of the $\nue$ and $\anue$ fluxes 
we expect that neutrino flavor oscillations between the neutrino-sphere 
and the
weak freeze-out radius  
will change the value of the electron 
fraction.

Such calculations were done by Qian et al. \cite{qian}.
They made a two flavor  
analysis of the matter-enhanced level crossing between $\nu_e $ and
 $\nu_\tau $ 
or $\nu_\mu $. They considered a mass spectrum in which $m_{\nu_{\tau,\mu}} 
>m_{\nu_e}$ so that there is resonance between the neutrinos 
only and not between the antineutrinos. 
As a result the more energetic $\nu_{\mu,\tau}$
 ($<E_{\nu_\mu}>$ = $<E_{\nu_\tau}>$ $\sim 25 MeV)$ get converted to $\nu_e $ 
($<E_{\nu_e}>\sim 11 MeV$) increasing the average energy of the
electron neutrinos.  
Since there is no transformation between the antineutrinos, the energy 
of the electron antineutrino remains the same( $<E_{\bar{\nu_e}}>\sim
16 MeV$).  
The mass of the $\nu_{\tau}$ (or $\nu_\mu)$ required to undergo MSW
resonance in the relevant region was shown to be between 1 and 100 eV which
is the right range for neutrinos to be the hot dark matter of the 
Universe. 
Finally they used the condition $Y_e < $  0.5 to put constraints on the 
mixing angle. 

There are several other hints of non-zero neutrino mass and
mixing coming from the solar neutrino experiments, the observation of
atmospheric neutrinos and from the recent LSND data.  

Four experiments (Homestake, Kamiokande, SAGE, Gallex) 
measuring the solar neutrino flux observed event 
rates significantly
smaller compared to the predictions of the Standard Solar Models. 
This contradiction constitutes the solar neutrino problem.
This can be explained by neutrino oscillations in vacuum for  
$\Delta m^2 \sim 0.615 \times 10^{-10} eV^2$ and $\sin^2{2\theta} 
\sim 0.864$ \cite{vac} or by the Mikheyev-Smirnov-Wolfenstein (MSW) resonant
flavor conversions \cite{msw}  for  ${\Delta m}^2 \sim  5.4 \times 10^{-6}
eV^2$ and $\sin^2{2\theta} \sim 7.9 \times 10^{-3}$ (non-adiabatic solution)
and ${\Delta m}^2 \sim 1.7 \times 10^{-5} eV^2$  and $\sin^2{2\theta} \sim 0.6
9$ (large-angle solution) \cite{msw2}. The preliminary results from the 
Super-Kamiokande confirm this deficit of solar neutrinos  
and favor the long wavelength vacuum oscillation solution \cite{sksolar}.

The atmospheric neutrino anomaly is the discrepancy in the measured
and expected values of the ratio of contained $\mu$-like and $e$-like 
events in the Kamiokande, IMB and Soudan experiments. This can be 
explained 
by $\nu_{\mu} - \nu_\tau$ or $\nu_\mu - \nu_e$ oscillations for
$\Delta m^2 \sim 10^{-2} eV^2$ and $\sin^2 2\theta \sim 1.0$ \cite{atm}.
A preliminary Super-Kamiokande data  has confirmed the
atmospheric neutrino problem for both sub-GeV and multi-GeV
neutrinos \cite{sk}.
Combining their result with the results from
other experiments, the allowed range 
for the $\nu_{\mu} - \nu_{\tau}$ mode is 
$4 \times 10^{-4} \leq \Delta m^2 \leq 5\times 10^{-3} eV^2$ at 90\% 
C.L. with $\sin^2 2\theta $ = 0.8 -1 \cite{mcg}. 
For the $\nu_{\mu} - \nu_e$ channel at 90\% C.L. the
allowed range is 
$10^{-3} - 7 \times 10^{-3} eV^2$ and $\sin^2 2\theta = 0.65 -1 $.  
If one combines the result of the CHOOZ experiment
\cite{chooz},
then the $\nu_{\mu} - \nu_e$ oscillation solution is ruled out at 90\%
C.L. A narrow range $6 \times 10^{-4} - 10^{-3} eV^2$ is allowed 
at 99\% C.L..\cite{mcg}. 

A third indication for a non-zero $\Delta m^2$ comes from the recent 
LSND data which gives the first evidence for $\bar{\nu}_\mu - {\bar\nu}_e$
\cite{lsnd} and $\nu_e - \nu_\mu$ oscillations using a laboratory
neutrino source.  This along with the non-observation of neutrino
oscillations in E776 at BNL \cite{e776} and in the reactor experiment 
Bugey \cite{bugey} 
indicates $\Delta m^2$ in the range 0.2 - 3 $eV^2$.  

In this paper we observe that for neutrino mass squared differences 
in the range $10^{-4} - 10^{-1} eV^2$ vacuum neutrino oscillations 
can take place
between the neutrino-sphere and weak freeze-out radius. 
Requiring that such oscillations are consistent with heavy element
nucleosynthesis from supernovae 
one can constrain the mixing of $\nu_e$ with $\nu_\mu$  
(or $\nu_\tau$) in this range. 
Since this is the relevant range for $\nu_\mu - \nu_e$ oscillation 
of atmospheric neutrinos, 
assuming such oscillations to be operative in supernovae one can compare
the allowed values of parameters.   
We first do a two-generation analysis keeping $\Delta m^2$ in this range.
We find that the condition $Y_e <0.5$ is satisfied for
all values of mixing angles. However using the condition
$Y_e < 0.45$ one can rule out the $\numu -\nue$ solution to 
the atmospheric neutrino anomaly 
in agreement with the CHOOZ result. Moreover since using the 
r-process constraint we can probe down to $\Delta m^2$ = $10^{-4}
eV^2$, the narrow range that was consistent with the CHOOZ data
can also be ruled out. 
If on the other hand we assume two-generation $\nu_e - \nu_\tau$
oscillations to be operative then this provides the only bound on
such mixing in this mass range. The earlier bounds on
$sin^2 2\theta_{e\tau}$ were for $\Delta m^2 > 80 - 100 eV^2$
\cite{ushida}. 

Next we go to the more realistic three-flavor picture. 
We consider the scenario where  $\Delta m_{12}^2 \sim 10^{-5} eV^2$ or
$10^{-11} eV^2$ 
and $\Delta m_{13}^2 \sim \Delta m_{23}^2 = \Delta m^2$
in the range  $10^{-4} - 10^{-1} eV^2$. 
This mass spectrum can simultaneously 
explain the solar and atmospheric neutrino data and has received
considerable interest in the recent past \cite{solatm}. 
We find that this scenario together with the 
CHOOZ constraint \cite{chooz} is consistent with $Y_e < 0.5$. 
However if we use the more stringent condition $Y_e < 0.45$  then
bounds can be given on the mixing parameters. For $\Delta m^2$  
$> 2\times 10^{-3} eV^2$ these bounds are weaker than
the CHOOZ bound but we can probe $\Delta m^2$ 
down to 
$10^{-4} eV^2$ which is not probed by any terrestrial experiment so far. 
The three-generation  scenario considered does not explain the LSND
data and  
to explain it as well one needs to introduce a sterile
neutrino \cite{sterile}. 

Another place in supernovae, where neutrino oscillations can have important
effect is in  
reviving the shock during the reheating epoch of the supernova \cite{fuller}. 
The result of neutrino oscillation is to enhance the rate of energy 
deposition by the neutrinos and hence produce a more energetic shock. 
We study the effect of vacuum oscillation of neutrinos 
in the delayed neutrino heating phase of  type II supernovae.  

The plan of the paper is as follows. In section 2 we discuss the r-process 
in the "hot bubble" and give the expression of $Y_e$.
In section 3 we first give 
the two-generation
vacuum oscillation formula and 
discuss the implications of such oscillations on the value of
$Y_e$ at the r-process epoch. We next give the three-generation
oscillation analysis. 
In the following section we discuss the effects of 
these  
oscillations 
on shock-reheating. Finally we present the conclusions.

\section{\bf Supernova r-process Nucleosynthesis}
Above the neutrino-sphere, the electron fraction $Y_e$ which is the second 
most important factor for r-process abundance calculations 
after the entropy per baryon, is determined by 
the competition between the rate of the reactions 
\begin{equation}
\nue+n \rightleftharpoons p+e^-
\label{1a}
\end{equation}
\begin{equation}
\anue+p \rightleftharpoons n+e^+
\label{1b}
\end{equation}
The expression for the value of $Y_e$ at freeze out is given by Qian et
al. \cite{qian} 
 as 
\begin{equation}
Y_e \approx 
{{1}\over{1+\lambda_{\anue p}/\lambda_{\nue n}}}
\label{ye}
\end{equation}
Where $\lambda_{\nue n}$ and $\lambda_{\anue p}$ are the reaction rates in 
(\ref{1a}) and (\ref{1b}).
The reaction rate $\lambda_{\nu N}$, where N can be either p or n 
is given by 
\begin{equation}
\lambda_{\nu N} \approx {{L_{\nu}}\over{4\pi r^2}} 
{{\int_0^{\infty} \sigma_{\nu N} (E) f_\nu (E) dE}\over{\int_0
^{\infty} Ef_\nu(E) dE}}  
\label{lamb}
\end{equation}
where $L_\nu$ is the neutrino luminosity (we consider identical luminosity 
for all the neutrino species), $\sigma_{\nu N}$ is the reaction cross-section  
and $f_\nu (E)$ is the normalised Fermi-Dirac 
distribution function with zero chemical potential 
\begin{equation}
f_\nu(E)={{1}\over{1.803 {T_\nu}^3}} {{E^2}\over{exp(E/T_\nu)+1}}
\end{equation}
where $T_\nu$ is the temperature of the particular neutrino concerned. The 
cross section is approximately given by \cite {fuller} 
\begin{equation}
\sigma_{\nu N} \approx 9.23\times 10^{-44}(E/MeV)^2 cm^2
\end{equation}
If we calculate $\lambda_{\nue n}$ and $\lambda_{\anue p}$ using eq 
(\ref{lamb}) then 
the expression for $Y_e$ becomes
\begin{equation}
Y_e \approx {{1}\over{1+T_{\anue}/T_{\nue}}}
\end{equation}
Typical values for the neutrino temperatures when r-process is operative are 
\cite {qian}, $T_{\nue}$ = 3.49 MeV, $T_{\anue}$ = 5.08 MeV and $T_{\numu}$ 
=7.94 MeV so that $Y_e \approx 0.41$. 
This being less than 0.5 neutron rich conditions are obtained 
in the hot bubble and r-process is possible. 

\section{\bf Neutrino Oscillations in Vacuum}

In the standard model of particle physics, neutrinos are assumed to be
massless.  
But there is no compelling reason for this assumption. If the neutrinos are 
indeed massive 
then in general 
the mass eigenstates will be different from 
the flavor eigenstates. The neutrinos are produced in the 
flavor eigenstates but propagate in their mass eigenstates. Because 
these bases are in general not identical, there will be mixing and neutrino 
flavor will not be conserved. 

\subsection{\bf Two Generation Analysis}
The two-generation conversion probability of an initial neutrino flavor 
$\nu_\alpha$ to a flavor $\nu_\beta$ after traveling a distance $L$ in vacuum
is given by 
\begin{equation}
P_{\nu_\alpha \nu_\beta} =\frac{1}{2} sin^2 2\theta ( 1 - cos 2 \pi L/\lambda)
\label{palphabeta}
\end{equation}
where $\lambda$ denotes the oscillation wavelength and can be
expressed as
\begin{equation}
\lambda = 2.5 \times 10^{-3} km \frac{E}{MeV} 
\frac{eV^2}{\Delta m^2}
\end{equation} 
The various limits of the equation (\ref{palphabeta}) are as follows,
 
\begin{itemize}
\item $\lambda >> L,
 P_{\nu_\alpha \nu_\beta} \rightarrow 0 $

\item $\lambda << L, 
P_{\nu_\alpha \nu_\beta} \rightarrow \frac{1}{2} \sin^2 2\theta $
when averaged over the source or detector distances or energy.

\item $\lambda \sim L/2, 
P_{\nu_\alpha \nu_\beta}= sin^2 2\theta$ 
\end{itemize}
Oscillation effects are observable for 
$\lambda \sim L$. 
Now for us, as we will discuss later in this section, $L$
$\sim$ 50 -100 km. 
Then, assuming an average value of $E \sim 10 MeV$ one gets
the condition $\Delta m^2 \sim  10^{-4} eV^2$. 
For higher values of $\Delta m^2$ the oscillation effects are averaged out 
and there is a constant conversion depending on the mixing angle. Beyond
1 $eV^2$ the MSW resonance being within the weak freeze-out radius \cite{qian}
matter enhanced resonant flavor conversions take  place.

As a result of flavor oscillations the neutrino energy distribution function 
itself will change to (assuming two flavors)
\begin{equation}
f_{\nue}^{osc}(E) = P_{\nue\nue}f_{\nue}(E) + P_{\numu\nue}f_{\numu}(E)\\
\label{f21}
\end{equation}
\begin{equation}
f_{\anue}^{osc}(E) = P_{\anue\anue}f_{\anue}(E) + P_{\anumu\anue}f_{\anumu}(E)
\end{equation}
For the probabilities we use the expression given in (\ref{palphabeta}).
The $L$ here corresponds to the distance between the neutrino-sphere
and the week freeze-out radius. For the values of $L$ relevant for
r-process  
we use the results of the 20 $M_\odot$ supernova model  
given in ref \cite {ww}.
According to this the weak freeze-out occurs at about 0.5 MeV while 
the r-process sets in much later at temperatures below 0.26 MeV 
between $t_{pb} \approx$ 3 to 15s. 
From the temperature vs. radius curves given in \cite{ww} we get 
the following values for the radii at two different times relevant for
supernova r-process nucleosynthesis at T=0.5 MeV : 
\begin{center}
For $t_{pb} = 3s,  r_{\small {WFO}} \approx 100 km $\\
$t_{pb} = 12s, r_{\small {WFO}} \approx 55 km $.
\end{center}
The position of the neutrino-sphere is $\sim 10km$.
Thus the $L$ relevant for us is $\sim$ 90 km at $t_{pb} = 3s$ and $\sim$
45 km at
$t_{pb} = 12s$. 

We find that for ${\Delta m}^2$ in the range from $10^{-4}
- 10^{-1} eV^2$ 
the value of $Y_e$ stays less than
0.5 for all values of mixing angles. 
For r-process to be possible the conditions should be 
neutron rich, that is, $Y_e <$ 0.5. But it is seen that for most r-process 
nuclei, the calculated abundances matches their solar abundances only for 
$Y_e <$ 0.45 \cite {ww,fcnc}. In Fig. 1 we give the contour plot for 
$Y_e$ = 0.45 for two different values of $L$. 
The region to the left of the lines are allowed by r-process (see
figure caption for details). 
The oscillation channel relevant 
is $\numu-\nue$ (or $\nu_{\tau} - \nu_e$) and the mass range that we study is 
$10^{-4}-10^{-1} eV^2$.
For comparison we also plot in the same figure the exclusion plot given 
by the CHOOZ experiment \cite {chooz} as well as the region allowed from 
the atmospheric neutrino data for the $\nu_\mu - \nu_e$ mode \cite{mcg}. 
For $\Delta m^2 \geq  2 \times 10^{-3} eV^2$ 
we get the constraint
$sin^2 2\theta_{e\mu} \leq 0.26$.
This is weaker than the 
CHOOZ result but nevertheless rules out the $\numu-\nue$ solution
to the atmospheric neutrino problem.  
Furthermore using r-process we can give bounds on the
mixing angle upto $\Delta m^2 = 10^{-4} eV^2$ and our results show
that even in the mass range $10^{-4} - 10^{-3} eV^2$ the condition $Y_e <
0.45$ is inconsistent with the $\numu - \nue$ solution of atmospheric
neutrino anomaly. Thus the narrow range which  was
allowed by the CHOOZ data at 
99\% C.L. is  ruled out by  r-process. 
The above conclusions are based on the assumption that the relevant
oscillation mode is $\nu_\mu - \nu_e$. If on the other hand we take
$\nu_e - \nu_\tau$ oscillations to be operative then this bounds will
apply to $sin^2 2\theta_{e \tau}$. In that case this provides the only
bound on $\nu_e -\nu_\tau$ mixing in this mass range \cite{ushida2} .

The equation (\ref{palphabeta}) is derived assuming that neutrinos 
can be represented by plane waves.  
But a correct quantum mechanical treatment of neutrino oscillation 
should describe them in terms of wave packets \cite{kim}. 
Then for extremely
relativistic neutrinos the two-flavor probability 
(\ref{palphabeta}) gets modified as \cite{kim}
\begin{equation}
P_{\nu_\alpha \nu_\beta} =\frac{1}{2} sin^2 2\theta 
( 1 - cos {\frac{2 \pi L}{\lambda}} e^{-L^2/L^2_{coh}})
\label{p2}
\end{equation}
where 
\begin{equation}
L_{coh} = 4 \sqrt{2} \sigma_x (E/\Delta m^2)
\label{lcoh} 
\end{equation}
where $\sigma_{x}$ denotes the spread of the wave packet. 
If $L << L_{coh}$, $e^{(-L^2/{L^2}_{coh})} \rightarrow 1$ 
one obtains  eq. (\ref{palphabeta}).  
If on the other hand $L >> L_{coh}$ 
$P_{\nu_\alpha \nu_\beta}= 0.5 sin^2 2\theta$.  
Thus in this case oscillations don't take place but there 
is a constant
conversion probability similar to the $\lambda << L$ case of the plane 
wave approximation. 
The important feature which comes out of the wave packet 
treatment is the coherence length $L_{coh}$. Beyond this
the separation between the wave packets 
associated with the propagating mass eigenstates becomes large enough
to cause any interference. Thus the oscillation can take place only
if 
the distance $ L \leq L_{coh}$. Otherwise there is a constant
conversion depending on the mixing angle.
For supernova neutrinos emitted from the
neutrino-sphere $\sigma_{x} \approx 10^{-9}$ cm 
\cite{kim} so that for 
$\Delta m^2$ in the range $10^{-4} - 10^{-1} eV^2$  the coherence
length varies from 5.656 $\times 10^{4} - 56.56$ km for   
$E \sim 10 ~MeV$.  
Thus in the range $10^{-4} - 10^{-2} eV^2$, $L_{coh} >> L$ and 
the plane wave approximation is valid. 
For $\Delta m^2 = 0.1 eV^2$ at low energies (upto 10 MeV) 
$L {\stackrel{<}{\sim}}  L_{coh}$  and the wave packet corrections
are important in principle. However because of the nature of the
Fermi-Dirac distribution function very few neutrinos of low energies are
present and thus 
this does not make any difference in practice.  
We have checked numerically that use of the wave packet formula (\ref{p2}) 
does not change our results. 

\subsection{Three Generation Analysis}
In this section we study the impact of 
three-generation vacuum oscillation of 
neutrinos on the freeze-out value of
$Y_e$.  
We fix one of the mass 
squared differences $\Delta m_{12}^2 \sim 10^{-5}$ or $10^{-11} eV^2$, 
corresponding to solar neutrinos and take the other two equal to each 
other i.e. $\Delta m_{23}^2\approx \Delta m_{13}^2$, each being in the 
range $10^{-4} - 10^{-1} eV^2$, suitable for atmospheric neutrino
oscillations. 
The general expression for the probability that an initial $\nu_{\alpha}$ 
of energy E gets converted to a $\nu_{\beta}$ after traveling a 
distance $L$ 
in vacuum is,
\begin{equation} 
P_{\nu_{\alpha}\nu_{\beta}}=\delta_{\alpha\beta}-4\sum_{j>i}
U_{\alpha i}  
U_{\beta i} U_{\alpha j} U_{\beta j} Sin^2 {{\pi L}\over{\lambda_{ij}}}
\label{p3}
\end{equation}
where, $\alpha = e, \mu, \tau$ and $i,j = 1, 2, 3$  
\begin{itemize}
\item $\lambda_{ij} = 2.5 \times 10^{-3} km \frac{E}{MeV} 
\frac{eV^2} { {\Delta m^2}_{ij} }$
\item $\Delta m_{ij}^2=m_j^2-m_i^2$ 
\end{itemize}
We  neglect CP violation in the lepton sector so that 
probabilities of neutrinos and antineutrinos are same. 
For the mass spectrum under consideration 
$\Delta m_{12}^2\sim 
10^{-5} eV^2$ or $10^{-11} eV^2$, so that $sin^2 \frac{\pi L}{\lambda_{12}}
\rightarrow 0$ and $\Delta m_{13}^2\approx \Delta m_{23}^2$. 
Thus one mass scale dominance (OMSD) applies. 
Using the orthogonality of the mixing matrix U  
the various probabilities relevant for us are 
\begin{equation}
P_{\nue\nue}= P_{\anue\anue} =
1-4({|U_{e3}|}^2(1 - {|U_{e3}|}^2))sin^2 \frac{\pi L}
{\lambda_{13}}
\label{pr3}
\end{equation}
\begin{eqnarray}
P_{\numu\nue}+P_{\nutau\nue}&=&P_{\anumu\anue}+P_{\anutau\anue}
\nonumber \\
                            &=&4({|U_{e3}|}^2(1-{|U_{e3}|}^2))
sin^2\frac{\pi L}{\lambda_{13}}    
\label{pr4}
\end{eqnarray}
We note that since the one mass scale
dominance holds good the probabilities are functions of only 
one mass squared difference. Secondly since the energy 
spectra of the $\numu$
and $\nutau$ are identical, the conversion probablities to $\nu_\mu$
and $\nu_\tau$  always appear as  
$P_{\numu\nue} + P_{\numu\nutau}$ 
and the probabilities depend on only
$U_{e3}$. 
For 3 flavor neutrino oscillations the 
neutrino distribution function becomes
\begin{equation}
f_{\nue}^{osc}(E)= P_{\nue\nue}f_{\nue}(E) + (P_{\numu\nue}+
P_{\nutau\nue})f_{\numu}(E)\\
\label{f31}
\end{equation}
\begin{equation}
f_{\anue}^{osc}(E)= P_{\anue\anue}f_{\anue}(E) + (P_{\anumu\anue}+
P_{\anutau\anue}) f_{\anumu}(E)
\label{f32}
\end{equation}
From expression (\ref{pr3}) we see that in the limit of OMSD 3 flavor 
oscillation reduces effectively to 2 flavor case with 
$sin^2 2\theta$ for the 
2 generation case replaced by the corresponding factor 
$4{|U_{e3}|}^2(1-{|U_{e3}|}^2)$. Therefore just as in 
the case of 2 flavor 
oscillations $Y_e <$ 0.5 constraint is never violated 
in our model and hence 
if this is the only constraint on the value of $Y_e$ then 
r-process is compatible with the solar neutrino data, 
the terrestrial accelerator-reactor data (excluding LSND) as well 
as data coming from atmospheric neutrino anomaly. 
However if we use  
$Y_e <$ 0.45  we will get a contour in the mass-mixing angle 
plane same as Fig. 1 with the relevant mass and 
mixing parameters being $\Delta m_{13}^2$ and 
$4{|U_{e3}|}^2(1-{|U_{e3}|}^2)$ 
instead of $\Delta m^2$ and $sin^2 2\theta$ respectively.  
It turns out that in the three generation 
framework that we have chosen, these are also the 
parameters relevant for the CHOOZ experiment \cite{guinti}.
We have compared the CHOOZ result with our result in Fig. 1. In 
the range $2 \times 10^{-3}-10^{-2} eV^2$ CHOOZ gives a slightly 
stronger constraint 
on $U_{e3}$ than r-process but there is an overall 
agreement between the two. 
Our results like CHOOZ \cite{guinti}  also imply a decoupling of solar
and atmospheric neutrino 
oscillations into separate two-genaration pictures for $\Delta m_{13}^2
> 10^{-3} eV^2$. 
In the range $10^{-4}\leq \Delta m_{13}^2 \leq 10^{-3} eV^2$ only
r-process can give constraints on the mixing parameter $U_{e3}$.

\section{\bf Shock Reheating}

We finally study the effect of vacuum neutrino oscillations on supernova 
dynamics. The prompt shock generally stalls at a radius of $\sim$ 100 km 
mainly due to energy loss from the shock radiated in neutrinos, 
energy loss 
due to dissociation of iron-group nuclei and due to accretion. In the 
delayed neutrino heating scenario \cite{bw} the matter behind the shock is 
heated by the energy deposited by the neutrinos 
through the capture processes 
(\ref{1a}) and (\ref{1b}), leading to an explosion.

If neutrinos have mass they will oscillate leading to the conversion of more 
energetic muon and tau neutrinos and antineutrinos to 
electron type neutrinos 
and antineutrinos, which results in enhanced neutrino energy deposition 
resulting in a more energetic shock. The effect of MSW conversion
\cite{fuller},  
resonant spin flavor precession \cite{rsfp} and resonant conversion of 
massless neutrinos due to flavor-changing neutral current interaction 
\cite{fcnc} have been considered before. In this work we examine the 
effect of vacuum neutrino oscillations on the rate of 
shock reheating and 
compare it to the corresponding values obtained earlier.

The rate at which energy is deposited by the neutrino 
capture on nucleons 
is given by \cite{bw}
\begin{equation}
\dot E\approx \frac{1}{4 \pi R_m ^2}[K_n(T_\nue)L_\nue + K_p(T_\anue)L_\anue]
\label{dotE}
\end{equation}
where $L_\nue$ and $L_\anue$ are the total $\nue$ and $\anue$ 
liminosities, 
$R_m$ is the radius and $K_n$ and $K_p$ are the neutrino absorption 
coefficients due to the reactions (\ref{1a}) and (\ref{1b})
\begin{equation}
K_\alpha=N_A Y_\alpha \langle \sigma(E_\nu) \rangle , \alpha=p,n
\end{equation}
where $N_A$ is the Avagadro's number, $Y_\alpha$ is the nucleon number per 
baryon and $\langle \sigma(E_\nu) \rangle$ is the reaction cross-section 
\cite{fuller} averaged over the neutrino spectrum, e.i.
\begin{equation}
\langle \sigma(E_\nu) \rangle = (9.23 \times 10^{-44})\int\limits_{0}^{\infty} 
f_\nu (E_\nu) {E_\nu}^2 dE_\nu
\end{equation}
In the expression (\ref{dotE}) we have neglected the neutrino energy loss 
term from $e^{\pm}$ capture for simplicity \cite{fuller}.

The luminosities of the different neutrino species are found to 
be almost equal in supernova models. 
For simplicity we consider identical energy spectrum 
for the $\nue$ and $\anue$ and the matter to be composed entirely 
of free nucleons. Then the  heating rate becomes 
\begin{equation}
\dot E \approx \frac{L_\nue}{4\pi R_m ^2}N_A(9.23\times 10^{-44}) 
\int\limits_{0}^{\infty} f_\nue (E_\nu) {E_\nu}^2 dE_\nu
\end{equation}
If one takes into account the neutrino flavor oscillations then 
the rate of heating is increased by
\begin{equation}
\frac{\dot {E_{osc}}}{\dot E}=\frac{\int\limits_{0}^{\infty} 
f_\nue^{osc} (E_\nu) {E_\nu}^2 dE_\nu}{\int\limits_{0}^{\infty}
f_\nue (E_\nu) {E_\nu}^2 dE_\nu} 
\end{equation}
where $f_\nue^{osc}$ is given by  
(\ref{f31}).

We perform our calculations for $\langle E_\nue \rangle = 
\langle E_\anue \rangle 
\approx$ 15 MeV and $\langle E_\numu \rangle \approx$ 21 MeV 
\cite{fcnc} and with $L \sim $ 100 km. 
We  do our calculations for $\Delta m_{12}^2 \sim 
10^{-5}$ or $10^{-11} eV^2$ and 
$\Delta m_{23}^2 \approx \Delta m_{13}^2 \sim 10^{-4} - 10^{-1}  eV^2$.
For $\Delta m_{13}^2$ in the range $10^{-3} - 10^{-1} eV^2$ 
we use the 
CHOOZ limiting value of 0.18, which is also consistent with r-process  
constraints, for the mixing parameter
$4{|U_{e3}|}^2(1-{|U_{e3}|}^2)$.
For $\Delta m_{13}^2 = 10^{-4} eV^2$ there is no constraint from
CHOOZ and for this we use the maximum possible value of the
mixing parameter allowed by r-process. 
The value of ${\rm \dot E_{osc}/\dot E}$ obtained is presented in Table
1.
Since above $2 \times 10^{-3} eV^2$ one gets average oscillations the
heating rate is independent of the value of $\Delta m^2$. 
It is seen that the increase in heating obtained is not much.
The reason for such low value 
of $\dot {E_{osc}}/\dot E$ can be traced to the very 
small value of $U_{e3}$ 
allowed by CHOOZ and r-process. 
For $\Delta m^2 = 10^{-4} eV^2$ although the mixing parameter 
can be large the factor $\sin^2 \pi L/\lambda$ itself becomes
very small
and there is very little effect of oscillations 
on the heating rate.

\section{Conclusion}

In this paper we have studied the effect of vacuum neutrino oscillations 
on supernova nucleosynthesis and dynamics. 
The mass range that we can probe is from $10^{-4} - 10^{-1} eV^2$. 
The position of the MSW resonance for such mass
squared values is beyond the weak freeze-out radius.  
We have studied the effect of such oscillations 
on the freeze-out value of $Y_e$ during the r-process nucleosynthesis
epoch and use the condition $Y_e < 0.45$ as the criterion for 
a successful r-process \cite{fcnc,ww} to  
give constraints on the  mixing between $\nu_e$ and $\nu_\mu$ 
(or $\nu_\tau$) from a two generation analysis. 
If we assume two-generation $\nu_e - \nu_\mu$ oscillations 
to be operative then r-process rules out this mode as a solution 
to the atmospheric neutrino anomaly, a result consistent with CHOOZ. 
Even the small range that is allowed by CHOOZ at 99\% C.L. is not 
allowed by r-process.
On the other hand if we assume two-genaration $\nu_e - \nu_\tau$
oscillations to be operative then this provides the only bound
on $\sin^2 2\theta_{e \tau}$ in this mass range. 

We next go to the more realistic
three-generation picture and take 
a scenario of mass  and mixing angles
which simultaneously explains (i) the solar neutrino problem and 
(ii) the atmospheric neutrino puzzle.
The length scales involved are such that the one-mass scale
dominance limit applies and the probabilities are functions of only
one $\Delta m^2$ and one gets an effective two-generation picture with 
$\Delta m^2$ replaced by 
$\Delta m_{13} \approx \Delta m_{23}$ and $sin^2 2\theta$ replaced 
by $4 U_{e3}^2 (1 - U_{e3}^2)$. In the three-generation scheme under
consideration these are also the parameters relevant for CHOOZ
and a comparison of the two shows that 
our constraints on the mixing angles 
for the mass range $2 \times 10^{-3}-10^{-1} eV^2$ are somewhat weaker. 
But for the mass range $10^{-4}-10^{-3} eV^2$ only r- 
process can give bounds on the mixing parameters
$\theta_{e\mu}$($\theta_{e\tau})$ in two-
generations and $U_{e3}$ in three-generations.

We also calculate the increase in the shock-reheating obtained via 
vacuum neutrino oscillations. 
In three-generation framework with the 
CHOOZ constraint we obtain $\dot {E_{osc}}/\dot E \approx$ 1.1.
Since $U_{e3}$ is very low the mixing is 
very small resulting in very little change in the heating rate. 

\vskip 20pt
\parindent 0pt

The authors would like to thank Kamales Kar for many helpful 
discussions and encouragement. 

\newpage
\begin{description}
\begin{center}
\item{Table 1.} The ratio of the heating rate with and without 
oscillations for different mass and mixing parameters.

\end{center}  
\end{description}
\[
\begin{array}{|c|c|c|c|c|} \hline
{\Delta m_{13}^2} & 
{\rm 4|U_{e3}|^2(1-|U_{e3}|^2)} & {\rm \dot E_{osc}/\dot E} 
\\ \hline
{10^{-1} eV^2} & {0.18} & {1.087}  \\ \hline
{10^{-2} eV^2} & {0.18} & {1.087}  \\ \hline
{10^{-3} eV^2} & {0.18} & {1.073} \\ \hline
{10^{-4} eV^2} & {0.90} & {1.067}\\ \hline
\end{array}
\]
\vskip 50pt
\begin{center}
{\bf Figure Caption}
\end{center}
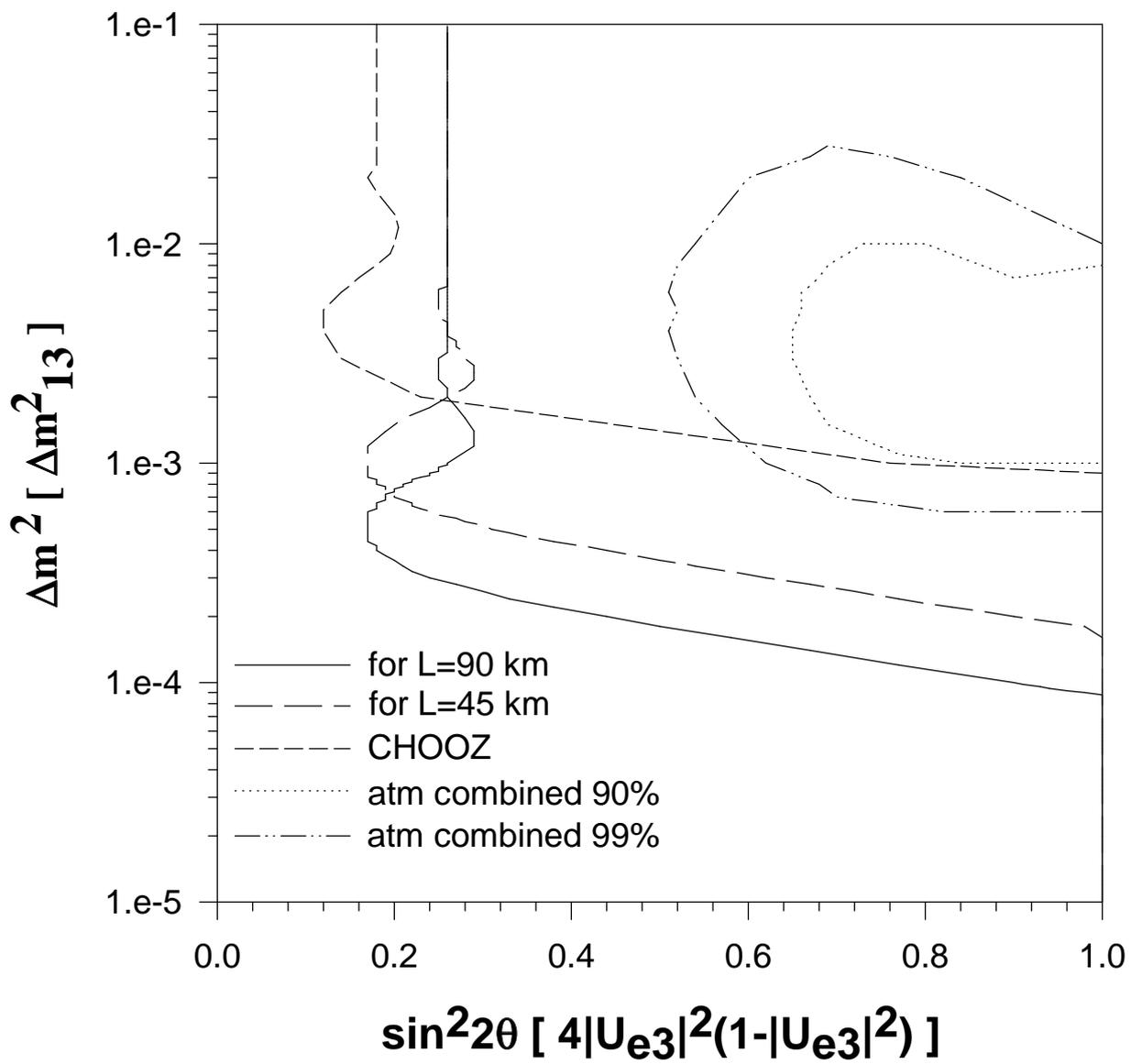
Fig 1. The $Y_e$ = 0.45 contours at two different post bounce times 
3s ( L=90 km) and 12s ( L=45 km). The region to the left of the lines 
is allowed by r-process. The mass and mixing parameters relevant 
for three-flavor mixing are given within brackets.
Also shown are the allowed parameters of the 
$\numu - \nue$ oscillation mode for the atmospheric neutrino data for 
all experiments combined at 90\% and 99\% C.L. and the exclusion plot 
from the CHOOZ experiment.

\end{document}